\documentclass[aps,prl,twocolumn]{revtex4}
\usepackage{graphicx}

\begin{document}


\title{Enhancing the Superconducting Transition Temperature due to Strong-Coupling Effect under Antiferromagnetic Spin Fluctuations in CeRh$_{1-x}$Ir$_{x}$In$_5$ : $^{115}$In-NQR Study}

\author{Shinji~Kawasaki$^1$}%
\author{Mitsuharu~Yashima$^1$}
\author{Yoichi Mugino$^1$}
\author{Hidekazu~Mukuda$^1$}%
\author{Yoshio~Kitaoka$^1$}%
\author{Hiroaki~Shishido$^2$}
\author{Yoshichika~\=Onuki$^2$}

\affiliation{$^1$Department of Materials Engineering Science, Graduate School of Engineering Science, Osaka University, Toyonaka, Osaka 560-8531, Japan\\$^2$Department of Physics, Graduate School of Science, Osaka University, Toyonaka, Osaka 560-0043, Japan}%


\date{\today}

\begin{abstract}
We report on systematic evolutions of antiferromagnetic (AFM) spin fluctuations and unconventional superconductivity (SC) in heavy-fermion (HF) compounds CeRh$_{1-x}$Ir$_{x}$In$_5$ via $^{115}$In nuclear-quadrupole-resonance (NQR) experiment. The measurements of nuclear spin-lattice relaxation rate $1/T_1$ have revealed the marked development of AFM spin fluctuations as a consequence of approaching an AFM ordered state with increasing Rh content. Concomitantly the superconducting transition temperature $T_{\rm c}$ and the energy gap $\Delta_0$ increase drastically from $T_{\rm c} = 0.4$ K and $2\Delta_0/k_{\rm B}T_{\rm c} = 5$ in CeIrIn$_5$ up to $T_{\rm c} = 1.2$ K and $2\Delta_0/k_{\rm B}T_{\rm c} = 8.3$ in CeRh$_{0.3}$Ir$_{0.7}$In$_5$, respectively. The present work suggests that the AFM spin fluctuations in close proximity to the AFM quantum critical point are indeed responsible for the onset of strong-coupling unconventional SC with the line node in the gap function in HF compounds. 
\end{abstract}

\pacs{}


\maketitle 
Unconventional superconductivity (SC) observed around antiferromagnetic (AFM) quantum critical point (QCP) has attracted broad interest and attention as one of the new topical issues in condensed-matter physics. For example, on the verge of magnetic order in heavy-fermion (HF) systems such as CeCu$_2$Si$_2$\cite{steglich}, CeRh$_2$Si$_2$\cite{Movshovic96}, CeIn$_3$, CePd$_2$Si$_2$\cite{Mathur98,Grosche01,Walker97}, and CeRhIn$_5$\cite{Hegger}, the magnetically soft electron liquid can mediate spin-dependent attractive interactions between the charge carriers\cite{Mathur98}. These findings suggest that the mechanism forming Cooper pairs can be magnetic in origin. However, there is no convincing experimental proof that the magnetic fluctuations in close proximity to the AFM-QCP  are a subtle origin for the onset of unconventional SC.

Recently, on the one hand, superconducting two domes have been reported as the function of pressure on CeCu$_2$(Si$_{1-x}$Ge$_{x}$)$_2$ \cite{Yuan}, being a topical issue on the origin of the SC in HF compounds. One dome (SC1) is formed around the AFM-QCP, whereas  another one (SC2) emerges under the HF state without any signature for AFM spin fluctuations because the system is far from the AFM-QCP. Interestingly, a maximum $T_{\rm c}$ in SC2 as the function of pressure is higher than that in SC1 for CeCu$_2$(Si$_{1-x}$Ge$_{x}$)$_2$. Although a possible origin of SC2 is not yet known, a new type of pairing mechanism is suggested to mediate the Cooper pairs in HF systems besides AFM spin fluctuations. For instance, valence fluctuations of Ce ions  may be responsible for the onset of SC2 via the increase of hybridization between conduction electrons and Ce-4$f$ electrons \cite{Yuan,Onishi,Miyake}. Two superconducting domes are also suggested in CeRh$_{1-x}$Ir$_x$In$_5$ \cite{Pagliuso,Nicklas}. The Rh substitution for Ir in the HF superconductor CeIrIn$_5$ decreases $T_{\rm c}$ down to $T_{\rm c} < 0.3$ K around $x$ = 0.9, but with a further increase of Rh substitution, the superconducting transition temperature $T_{\rm c}$ increases up to 1 K for CeRh$_{0.5}$Ir$_{0.5}$In$_5$ \cite{Pagliuso}. Note that the Rh substitution increases the $c/a$ ratio \cite{Pagliuso}. For 0.35 $\le$ x $\le$ 0.5, the uniform coexistence of antiferromagnetism and SC was suggested by the previous nuclear-quadrupole-resonance (NQR) measurement by Zheng $et$ $al$.\cite{Zheng2} This coexistent phase is characterized by the gapless nature of its SC \cite{Zheng}. Such the coexistence of antiferromagnetism and gapless SC was reported to take place in CeRhIn$_5$ under pressure \cite{ShinjiRh}. Notably, the application of pressure also makes $T_{\rm c}$ increase up to 1 K around $P$$\sim$3 GPa in CeIrIn$_5$ \cite{Borth,Muramatsu}. Our previous NQR study showed that the unconventional SC in CeIrIn$_5$ under pressure is realized in the HF state without AFM spin fluctuations \cite{Shinji}. In this context, it is likely that there are two mechanisms to increase $T_{\rm c}$ in CeRh$_{1-x}$Ir$_x$In$_5$.  However, it is still unknown why $T_{\rm c} = 0.8$ K in CeRh$_{0.5}$Ir$_{0.5}$In$_5$ becomes larger than  $T_{\rm c} = 0.4$ K from CeIrIn$_5$, although impurity effect due to the Rh substitution for Ir site is believed to reduce $T_{\rm c}$. 

In this Letter, we report on systematic evolutions of superconducting characteristics and AFM spin fluctuations in CeRh$_{1-x}$Ir$_x$In$_5$ through the NQR measurements of nuclear spin-lattice relaxation time $T_1$ under zero field ($H=0$).  
Present results suggest that the AFM spin fluctuations in close proximity to the AFM-QCP are indeed responsible for the strong-coupling SC with the line node in the gap function in CeRh$_{1-x}$Ir$_x$In$_5$.  A reason why $T_{\rm c}$ has the minimum value of $T_{\rm c}^{\rm min}$ = 0.35 K at $x$ = 0.9 is proposed as due to the presence of two superconducting domes in the phase diagrams as the function of lattice density under the Rh substitution for Ir in  CeRh$_{1-x}$Ir$_x$In$_5$ and under the pressure for CeIrIn$_5$.

Single crystals of CeRh$_{1-x}$Ir$_{x}$In$_5$ ($x$ = 0.6 $-$ 0.9) were grown by the self-flux method and moderately crushed into grains in order to allow for radio-frequency pulses to penetrate easily into samples. Here the grain's size were kept larger than 100 $\mu$m to avoid possible damage into the sample quality. CeIr(Rh)In$_5$ consists of alternating layers of CeIn$_3$ and Ir(Rh)In$_2$ and hence has two inequivalent $^{115}$In sites per one unit cell. The $^{115}$In-NQR measurements were made at the In(1) site which is located on the top and bottom faces in the Ce-In layer of the tetragonal unit cell and hence affected sensitively by the Ce-4$f$ derived superconducting and magnetic phenomena. The $^{115}$In-NQR measurement was made by a conventional saturation-recovery method. The $T_1$ was measured at the transition of 2$\nu_{\rm Q}$ ($\pm 3/2\leftrightarrow \pm 5/2$) down to 0.05 K, using a $^3$He-$^4$He dilution refrigerator.  
\begin{figure}[h]
\centering
\includegraphics[width=6cm]{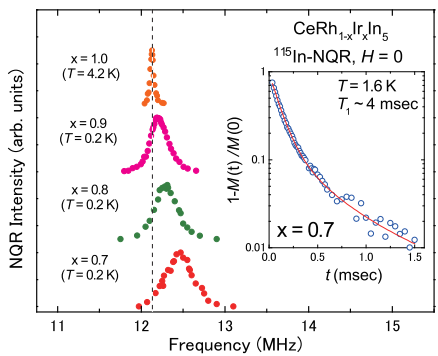}
\caption[]{\footnotesize (color online) $^{115}$In-NQR spectra (2$\nu_{\rm Q}$ transition) of the In(1) site at $x$ = 0.7, 0.8, and 0.9 in CeRh$_{1-x}$Ir$_{x}$In$_5$ along with the spectrum of CeIrIn$_5$ cited from Ref.\cite{Zheng}. The dotted line indicates the peak position for $x$ = 1.0. Inset shows a recovery curve of  nuclear magnetization of the In(1) site at 1.6 K for $x$ = 0.7. A solid curve indicates the theoretical one (see text).}
\end{figure}
 

Figure 1 shows the $^{115}$In-NQR spectra (2$\nu_{\rm Q}$ transition) at 0.2 K for $x$ = 0.9, 0.8, and 0.7 in CeRh$_{1-x}$Ir$_{x}$In$_5$ along with the spectrum at 4.2 K for CeIrIn$_5$ ($x$ = 1.0) reported previously \cite{Zheng}. The Rh substitution for Ir makes the NQR spectral width increase associated with an inhomogeneous distribution in the electric field gradient at the In(1) site. However, the NQR frequency $\nu_{\rm Q}(x)$ increases linearly with an increase of Rh content, which suggests a progressive change in the lattice parameter due to the Rh substitution. The present $x$ dependence of $\nu_{\rm Q}$ is consistent with the previous report \cite{Zheng}.  Figure 1 inset indicates a typical NQR recovery curve of the nuclear magnetization of the In(1) site at 1.6 K for $x$ = 0.7, which is well fitted by the theoretical curve (solid line) \cite{InT1} with a single $T_1$ component irrespective of Rh content. This result assures that even though Rh is substituted for Ir, the electronic state is uniquely determined for all samples, allowing us to extract a systematic evolution in magnetic and superconducting characteristics as the function of Rh substitution. 

\begin{figure}[h]
\centering
\includegraphics[width=5.5cm]{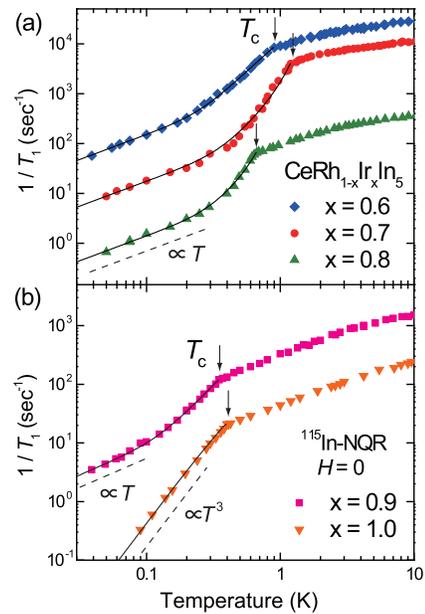}
\caption[]{\footnotesize (color online) $T$ dependence of $1/T_1$ at (a) $x$ = 0.6, 0.7, and 0.8 and (b)  $x$ = 0.9 and 1.0 in CeRh$_{1-x}$Ir$_{x}$In$_5$. The data for $x$ = 1.0 are taken from Ref.\cite{Zheng}.  The $1/T_1$ data at $x$ = 0.9, 0.7, and 0.6 are offset for clarity, where the respective raw data are multiplied by 5, 20, and 40 times. Arrows indicate each $T_{\rm c}$. Solid lines below $T_{\rm c}$ indicate the calculations based on the unconventional superconducting model (see text). Dashed lines indicate the relation of either $1/T_1\propto T^3$ or $1/T_1\propto T$, respectively. }

\end{figure}

\begin{figure}[h]
\centering
\includegraphics[width=6cm]{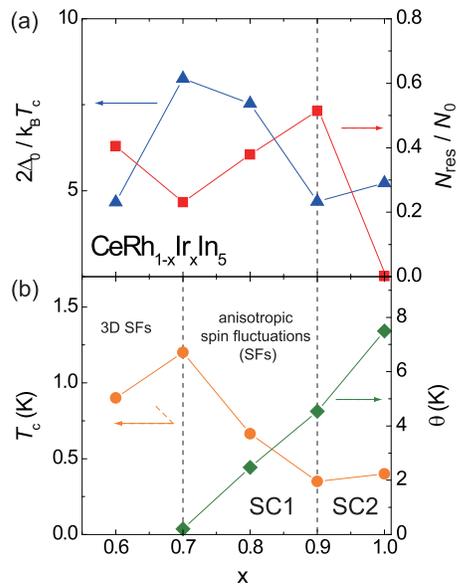}
\caption[]{\footnotesize (color online) (a) The Rh content dependence of $2\Delta_0$/k$_{\rm B}$$T_{\rm c}$ and $N_{\rm res}$/$N_0$. Lines are guide to the eye. (b) The $x$ dependence of $T_{\rm c}$ and $\theta$ obtained by the fitting of anisotropic AFM spin fluctuations model (see text). A dotted line at $x = 0.7$ denotes that the character of AFM spin fluctuations is changed from the isotropic to anisotropic ones, and it at $x = 0.9$ does that the character of SC is separated into SC1 and SC2 (see text). The error bars to determine these values in (a) and (b) are within the size of their symbol marks.}
\end{figure}

Figure 2 shows the temperature ($T$) dependence of $1/T_1$ at $x$ = 0.6 $-$ 0.9 in CeRh$_{1-x}$Ir$_{x}$In$_5$ along with the data for CeIrIn$_5$ ($x$ = 1.0) \cite{Zheng}. The steep decrease in $1/T_1$ denoted by down arrows is due to the onset of superconducting transition for all samples. Here, the absence of the coherence peak in $1/T_1$ just below $T_{\rm c}$ and the $T^3$ dependence for $x$ = 1.0 upon cooling give evidence for the unconventional nature of SC.  Note that $T_{\rm c}$ has the minimum value of $T_{\rm c}^{\rm min}$ = 0.35 K at $x$ = 0.9 although $T_{\rm c}$ increases as Rh content increases and has the maximum value of $T_{\rm c}^{\rm max}$ = 1.2 K at $x$ = 0.7.  It should be noted that $1/T_1$ tends to be proportional to the temperature well below $T_{\rm c}$. The most straightforward explanation for this relaxation behavior at low temperatures would be the presence of disorder that produces residual density of states (RDOS) remaining at the Fermi level ($E_{\rm F}$). By assuming the line node and the RDOS at $E_{\rm F}$ in a gap function with $\Delta(\theta)=\Delta_0\cos\theta$, we have tried to fit the $1/T_1$ data in the superconducting state to
\[
\frac{T_1(T_{\rm c})}{T_{1}}=\frac{2}{k_{\rm B}T} \int \left( \frac{N_{\rm S}(E)}{N_0} \right)^2 f(E) [1-f(E)] dE,
\]
where $N_{\rm S}(E)/N_0=E/\sqrt{E^2-\Delta^2}$ with $N_0$ being the DOS at $E_{\rm F}$ in the normal state and $f(E)$ is the Fermi-distribution function. From fittings shown by solid lines in Fig.~2, $T_{\rm c}$, 2$\Delta_0/k_{\rm B}T_{\rm c}$, and $N_{\rm res}$/$N_0$ are plotted against the Rh content in Figs. 3(a) and 3(b).  2$\Delta_0/k_{\rm B}T_{\rm c}$ increases markedly as well as $T_{\rm c}$, reaching $2\Delta_0/k_{\rm B}T_{\rm c}$ = 8.3 at $x$ = 0.7 which is larger than the weak-coupling BCS value of $2\Delta_0/k_{\rm B}T_{\rm c}$ = 3.53. 
A reason why $T_{\rm c}$ goes up regardless of the presence of disorder that produces $N_{\rm res}$ is this strong-coupling effect for forming the Cooper-pairs. As a result, $T_{\rm c}$ increases as the system approaches the AFM-QCP, nevertheless the impurity effect associated with the Rh substitution for Ir would be expected to make $T_{\rm c}$ reduce in general.
It may be because the Ce-In layer plays a dominant role for their SC\cite{Daniel} due to their anisotropic crystal and electric structures\cite{Shishido}.
On the one hand, it is noteworthy that the $x$ = 0.9 sample shows the largest value of $N_{\rm res}$/$N_0$ = 0.51 and the lowest values of $T_{\rm c}$ = 0.35 K with $2\Delta_0/k_{\rm B}T_{\rm c}$ = 4.68, although the narrower NQR linewidth is rather indicative of less disorder effect in the $x$ = 0.9 sample than others (see Fig.1).
These anomalous behaviors for the $x$ = 0.9 sample may be related to some superconducting critical phenomenon where the superconducting dome (SC1) at the vicinity of the AFM phase crosses over to another dome (SC2) emerging under the HF state without any trace for AFM spin fluctuations. 

\begin{figure}[h]
\centering
\includegraphics[width=5.7cm]{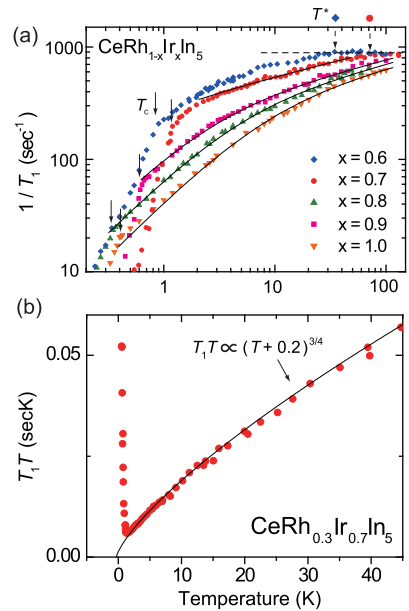}
\caption[]{\footnotesize (color online). (a) $T$ dependence of $1/T_1$ in both logarithmic scales at $x$ =0.6 $-$ 1.0. Solid curves are fittings based on the anisotropic AFM spin fluctuations model (see text). The data for $x$ = 1.0 are taken from Ref.\cite{Zheng}. Solid and dotted arrows indicate $T_{\rm c}(x)$ and $T^*$, respectively. Dashed line indicates the relations of $1/T_1$ = const. (b) $T_1T$ vs $T$ at $x$ = 0.7. Solid curve indicates the relation, $T_1T\propto (T + \theta)^{3/4}$ with $\theta$ = 0.2 $\pm$ 0.1 K.}
\end{figure}

Next, we deal with a systematic evolution of AFM spin fluctuations in the normal state in CeRh$_{1-x}$Ir$_x$In$_5$.
Figure 4(a) shows the $T$ dependence of $(1/T_1)_{4f}$ for $x$ = 0.6 $-$ 1.0. Above $T_{\rm c}$, the Ce-4$f$ electron contribution  $(1/T_1)_{4f}$ was evaluated by subtracting $(1/T_1T)_{\rm La}$ = 0.81 sec$^{-1}$K$^{-1}$
 of LaIrIn$_5$ \cite{Zheng} from the measured $(1/T_1)_{\rm Ce}$, namely, it is defined as $(1/T_1)_{4f}=(1/T_1)_{\rm Ce}-(1/T_1)_{\rm La}$. When the system is in close proximity to the AFM-QCP, the anisotropic AFM spin fluctuations model predicts a relation of $T_1T \propto 1/\chi_{\rm Q}(T) \propto (T + \theta)^{3/4}$ \cite{SCRreview,Lacroix,Kondo}. Here the staggered susceptibility with the AFM propagation vector ${\bf q = Q}$, $\chi_{\rm {\bf Q}}(T)$ follows a Curie-Weiss law. So, $\theta$ is one of measure to what extent is close to a QCP, i.e. $\theta$ = 0 means that the system is just on the QCP.
Actually, the $T$ dependences of $1/T_1$ in CeIrIn$_5$ and CeCoIn$_5$ are well fitted by this model \cite{Zheng,Yu}. Consistently with the previous examples, as seen from the tentative fittings shown by solid lines in Fig.4(a), the $T$ dependences of $1/T_1$ above $T_{\rm c}$ in CeRh$_{1-x}$Ir$_x$In$_5$ ($x$ $\ge$ 0.7) are also well explained  by this anisotropic AFM spin fluctuations model \cite{SCRreview,Lacroix,Kondo}.  In the case of CeRh$_{0.3}$Ir$_{0.7}$In$_5$, remarkably, the $T$ dependence of $T_1T$ above $T_{\rm c}$ allows us to deduce a value of $\theta$ = 0.2 $\pm$ 0.1 K, demonstrating that this compound is very close to the AFM-QCP as seen in Fig.4(b). As summarized in Fig.3(b), $\theta$ = 8, 4.5, 2.5, and 0.2 K are estimated for $x$ = 1.0, 0.9, 0.8, and 0.7, respectively.
Notably, the values of $\theta= 0.2\pm 0.1$ K and the maximum energy gap $2\Delta_0/k_{\rm B}T_{\rm c}$ = 8.3 with the highest $T_{\rm c}$ = 1.2 K in $x$ = 0.7 sample are comparable to $\theta\sim0$ K and $2\Delta_0/k_{\rm B}T_{\rm c}$ = 8.86 for CeCoIn$_5$\cite{Yu,Yashima}. In CeCoIn$_5$, the strong-coupling $d$$_{x^2-y^2}$ SC was experimentally and theoretically demonstrated to be mediated by the strong AFM spin fluctuations \cite{Kohori1_3,Izawa1_3,SIkeda1_3,HIkeda1_3,Yu,Yashima}. The magnetic and superconducting properties for the $x$ = 0.7 sample resemble those for CeCoIn$_5$. It suggests that this strong AFM spin fluctuations increase $T_{\rm c}$ in the present compounds, nevertheless the disorder is introduced by substituting Rh for Ir.  In this context,  magnetic fluctuations could be the mediators in the pairing, but they are not for the undoped sample and the case under the pressure. Here, note  that the $1/T_1$ at $x$ = 0.6 is not necessarily consistent with this model, but rather seems to be compared with the isotropic AFM spin-fluctuations model \cite{SCRreview}. Furthermore, at $x$ = 0.6, $1/T_1$ approaches a constant above $T^*$ = 25 K, showing that the system starts to enter a localized regime where AFM order may be stabilized \cite{Kohori,ShinjiT*}. Unfortunately, since substitution makes NQR spectrum broaden, tiny magnetic ordered moment expected at $x$ = 0.6 could not be detected in the present measurement. However, actually, the AFM order takes place for the $x$ = 0.5 sample \cite{Pagliuso,Zheng2} and the isotropic AFM spin-fluctuations model is applied to interpret the $T$ dependence of $1/T_1$ for CeRhIn$_5$ ($x$ = 0) \cite{Curro}. In this context, it is likely that the character of AFM spin fluctuations crosses over from the anisotropic to isotropic regime around the $x=0.6$ sample.

\begin{figure}[h]
\centering
\includegraphics[width=6.5cm]{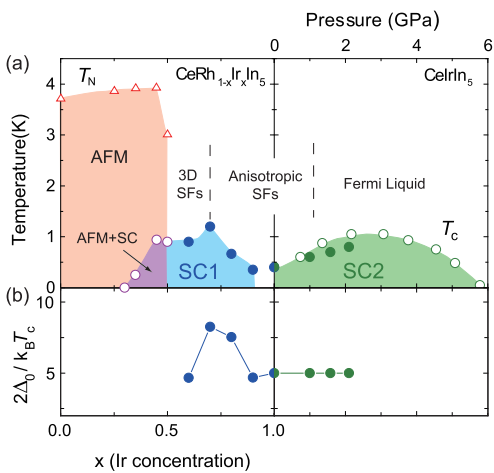}
\caption[]{\footnotesize (color online) (a) Phase diagram for CeIrIn$_5$ as the function of Ir concentration and the pressure. $T_{\rm c}$ (open circles) for CeIrIn$_5$ and $T_{\rm N}$ (open triangles) and $T_{\rm c}$ (open circles) for CeRh$_{1-x}$Ir$_x$In$_5$ are referred from refs.\cite{Zheng} and \cite{Muramatsu}, respectively. (b) $2\Delta_0/k_{\rm B}T_{\rm c}$ is plotted against the Ir content $x$ and the  pressure.}
\end{figure}

In conclusion, we suggest that the magnetic fluctuations in close proximity to the AFM-QCP are related to the strong-coupling Cooper-pairs formation in the $x = 0.7$ sample, leading to the highest $T_{\rm c}=1.2$ K and the largest energy gap $2\Delta_0/k_{\rm B}T_{\rm c}=8.3$. In Fig.5(a), the systematic evolutions of SC and AFM spin fluctuations are summarized as the function of the lattice density under the chemical substitution of Rh for Ir and the pressure. This phase diagram presents the rich variety of superconducting phenomena emerging in HF systems. (1) In $0.35 \le x \le 0.5$, the SC coexists with the AFM order on the microscopic level \cite{Zheng2}. This coexistent phase is characterized by the gapless nature of SC \cite{Zheng2}. (2) In $0.5 < x \le 0.9$, the SC1 dome is formed in close proximity to the AFM-QCP. (3) In $x >$ 0.9 and under the pressure, as the system is away from the AFM-QCP, the size of energy gap becomes smaller and remains constantly as shown in Fig.5(b), on the other hand, $T_{\rm c}$ goes up with the pressure. The SC2 dome emerges under the heavy-Fermi liquid state without any trace for AFM spin fluctuations \cite{Shinji}. 
In this context, the two superconducting domes have been thus evidenced in the Ce115 compounds, shedding new light on the superconducting phenomena emerging in HF systems. The present works, we believe, inspire a new view on the superconducting phenomena in strongly correlated electron systems in general.

This work was supported by the Grant-in-Aid for Creative Scientific Research (15GS0213) from the Ministry of Education, Culture, Sports, Science and Technology (MEXT) and the 21st Century COE Program (G18) by Japan Society of the Promotion of Science (JSPS)



\begin{references}
\bibitem{steglich} 
F.~Steglich $et$ $al$., Phys.~Rev.~Lett. {\bf 43}, 1892 (1979).
\bibitem{Movshovic96}
R.~Movshovich $et$ $al$., Phys.~Rev.~B {\bf 53}, 8241 (1996).
\bibitem{Mathur98}
N.~D.~Mathur $et$ $al$., Nature {\bf 394}, 39 (1998).

\bibitem{Grosche01}
F.~M.~Grosche $et$ $al$., J.~Phys.:~Condens.~Matter {\bf 13}, 2845 (2001). 

\bibitem{Walker97}
I.~R.~Walker $et$ $al$., Physica~C {\bf 282-287}, 303 (1997).


\bibitem{Hegger}
H. ~Hegger $et$ $al$., Phys. Rev. Lett. {\bf 84}, 4986 (2000).

\bibitem{Yuan}
H. Q. Yuan $et$ $al$., Science {\bf 302}, 2104 (2003).

\bibitem{Onishi} 
Y. Onishi and K. Miyake, J. Phys. Soc. Jpn. {\bf 69}, 3955 (2000).

\bibitem{Miyake}
K. Miyake and H. Maebashi, J. Phys. Soc. Jpn. {\bf 71}, 1007 (2002).

\bibitem{Pagliuso}
P. G. Pagliuso $et$ $al$., Phys. Rev. B {\bf 64}, 100503(R) (2001).

\bibitem{Nicklas}
M. Nicklas $et$ $al$., Phys. Rev. B, {\bf 70}, 020505(R) (2004).


\bibitem{Zheng2}
G.-q. Zheng $et$ $al$., Phys. Rev. B {\bf 70}, 014511 (2004).


\bibitem{Zheng} 
G.-q. ~Zheng $et$ $al$., Phys. Rev. Lett. {\bf 86}, 4664 (2001).
\bibitem{ShinjiRh} 
S.~Kawasaki $et$ $al$., Phys. Rev. Lett. {\bf 91}, 137001 (2003).



\bibitem{Muramatsu}
T.~Muramatsu $et$ $al$., Physica C {\bf 388-389}, 539 (2003).


\bibitem{Borth}
R.~Borth $et$ $al$., Physica B {\bf 312-313}, 136 (2002). 

\bibitem{Shinji}
S. Kawasaki $et$ $al$., Phys. Rev. Lett. {\bf 94}, 037007 (2005).



\bibitem{InT1}
D. E. MacLaughlin, J. D. Williamson, and J. Butterworth, Phys. Rev. B {\bf 4}, 60 (1971).

\bibitem{Daniel}
M. Daniel $et$ $al$., Phys. Rev. Lett. {\bf 95}, 016406 (2005).

\bibitem{Shishido}
H. Shishido, $et$ $al$., J. Phys. Soc. Jpn. {\bf 71}, 162 (2002). 


\bibitem{SCRreview} 
For review see, T. Moriya and K. Ueda, Adv. Phys. {\bf 49}, 555 (2000) and references therein. 

\bibitem{Lacroix}
C. Lacroix $et$ $al$., Phys. Rev. B {\bf 54}, 15178 (1996).

\bibitem{Kondo}
H. Kondo, J. Phys. Soc. Jpn {\bf 71}, 3011 (2002).


\bibitem{Yu} 
Y.~Kawasaki $et$ $al$., J. Phys. Soc. Jpn. {\bf 72}, 2308 (2003).

\bibitem{Yashima}
M. Yashima $et$ $al$., J. Phys. Soc. Jpn. {\bf 73}, 2073 (2004).


\bibitem{Kohori1_3} 
Y.~Kohori $et$ $al$., Phys.\ Rev.\ B {\bf 64}, 134526 (2001).

\bibitem{Izawa1_3} 
K. ~Izawa, $et$ $al$., Phys. Rev. Lett. {\bf 87}, 057002 (2001).

\bibitem{SIkeda1_3} 
S.~Ikeda $et$ $al$., J. Phys. Soc. Jpn. {\bf 70}, 2248 (2001).

\bibitem{HIkeda1_3}
H.~Ikeda, Y.~Nisikawa and K.~Yamada, J.\ Phys.: Condens.\ Matter {\bf 15}, S2241 (2003).

\bibitem{Kohori}
Y. Kohori $et$ $al$., Eur. Phys. J. B {\bf 18}, 601 (2000).

\bibitem{ShinjiT*}
S. Kawasaki $et$ $al$., Phys. Rev. B {\bf 65}, 020504(R) (2002). 

\bibitem{Curro}
N. J. Curro $et$ $al$., Phys. Rev. B. {\bf 62}, R6100 (2000).



\end{references}
\end{document}